\documentclass[12pt,a4paper]{article}
\usepackage{graphicx}
\usepackage{amsmath}

\begin{document}

\begin{center}
{\bf \Large New algorithm for the computation of~the~partition function for the
Ising model on~a~square lattice}\\[5mm]

{\em 
{\large
K.~Malarz$^{1,*}$,
M.S.~Magdo\'n-Maksymowicz$^2$,
A.Z.~Maksymowicz$^1$,
B.~Kawecka-Magiera$^1$
and~K.~Ku{\l}akowski$^1$
}

\bigskip

$^1$ Department of Applied Computer Science,
Faculty of Physics and Nuclear Techniques,
University of Mining and Metallurgy (AGH)\\
al. Mickiewicza 30, PL-30059 Krak\'ow, Poland\\

$^2$ Department of Mathematical Statistics, Agriculture University\\
al. Mickiewicza 21, PL-31120 Krak\'ow, Poland

\bigskip

E-mail:
$^*$malarz@agh.edu.pl\\[3mm]

March 17, 2003
}
\end{center}

\begin{abstract}
\noindent
A new and efficient algorithm is presented for the calculation of the partition
function in the $S=\pm 1$ Ising model. As an example, we use the algorithm to
obtain the thermal dependence of the magnetic spin susceptibility of an Ising
antiferromagnet for a $8\times 8$ square lattice with open boundary
conditions. The results agree qualitatively with the prediction of the Monte
Carlo simulations and with experimental data and they are better than the mean field 
approach results. For the $8\times 8$ lattice, the algorithm
reduces the computation time by nine orders of magnitude.
\end{abstract}


\section{Introduction}
A calculation of the partition function exactly in the thermodynamic limit is
equivalent to knowledge of all equilibrium properties of a given system.
This task is usually beyond our horizon, except for some selected cases which we
call ``trivial''.
The Ising model seems to be at the border of our possibilities since the beginning of the 20-th century.
Even for a finite system of $N$ spins our computational ability is limited by the number of the system
configurations, which is $2^N$.
Here we
describe a new algorithm, designed to improve the speed of the calculation of
the partition function for a finite Ising system.

	In the Ising model \cite{ising} only two spin states are possible, say ``up'' and ``down'' ($S_i=\pm 1$).  
The energy $E$ of a given spin configuration may be expressed in terms of the number $n$ of spins pointing ``up'' (say $S_i=+1$) and the number $k$ of anti-parallel bonds between the nearest neighbors $(S_iS_j=-1)$: 
\begin{equation} \label{eq_E}
E(n,k)=-J\sum_{<i,j>} S_iS_j -H\sum_i S_i=2J(k-L^2+L)-H(L^2-2n), 
\end{equation}
where $J$ denotes the exchange integral, $H$ is an external magnetic field, and
$L$ --- the linear size of the lattice.

	The partition function can be written as
\begin{equation}
\label{eq_Z}
Z=\sum_{n,k}\Omega(n,k)\cdot\exp[-\beta E(n,k)],
\end{equation}
where $\Omega(n,k)$ is the number of lattice configurations with given
numbers $n$ and $k$, $1/\beta=k_BT$, $k_B$ is a Boltzmann constant and $T$
is temperature. Then, the average value of any quantity $A$ may be calculated
as 
\begin{equation}
\label{eq_ave}
\langle A\rangle=Z^{-1}\sum_{n,k} A(n,k)\cdot\Omega(n,k)\cdot\exp[-\beta E(n,k)].
\end{equation}

The magnetic susceptibility per spin $\chi$ may be also expressed in the terms
of $n$, $k$ and $L$: 
\begin{equation}
\label{eq_chi}
\chi=\beta[\langle S_i^2\rangle-\langle S_i\rangle^2]=\beta[\langle
(2n-L^2)^2\rangle-\langle 2n-L^2\rangle^2],
\end{equation}
where the index $k$ enters through the averaging procedure.

	A computation of the partition function \eqref{eq_Z} requires an evaluation of the histogram $\Omega(n,k)$.
Tab. \ref{tab_cpu_time} shows the CPU time needed to check
the numbers $n$ and $k$ of $M$ configurations of a $8\times 8$
large lattice on SGI 2800 machine. A rough estimation of the CPU time necessary
for full investigation of all of $2^{64}\approx 10^{19}$ possible lattice
configurations gives the value larger than four millions years --- what makes
traditional/direct method practically useless.
\begin{table}
\begin{center}
\caption{
The CPU time necessary for investigation of $M$
different configurations of $8\times 8$ large 2D Ising lattice on SGI 2800
machine.} \label{tab_cpu_time} \vspace{2mm}
\begin{tabular}{lllllll}
\hline
$M$ & $10^5$ & $10^6$ & $10^7$ & $10^8$ & $10^9$ & $10^{10}$ \\
$t_{\text{CPU}}$ [sec]& 0.86 & 6.90 & 66.4 & 660 & 6648 & 65752 \\
\hline
\end{tabular}
\end{center}
\end{table}
\section{Calculations}

\subsection{The algorithm}
\label{sec_algorithm}

	An effective way of generation $\Omega_{8\times 8}(n,k)$ is
a successive merging of smaller lattices and their histograms, namely
$\Omega_{4\times 4}(n,k)$ and $\Omega_{8\times 4}(n,k)$. However, the
procedure requires storing information on the $\Omega$ dependence not only on
$n$ and $k$, but also on $b^r$ --- the state of the $r$-sites-long lattice
border.

\begin{figure}
\begin{center}
(a) \includegraphics[scale=.3]{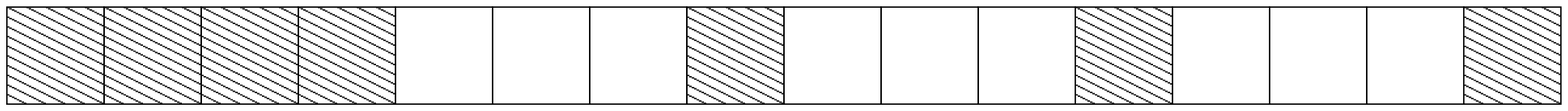}\\[2mm]
(b) \includegraphics[scale=.3]{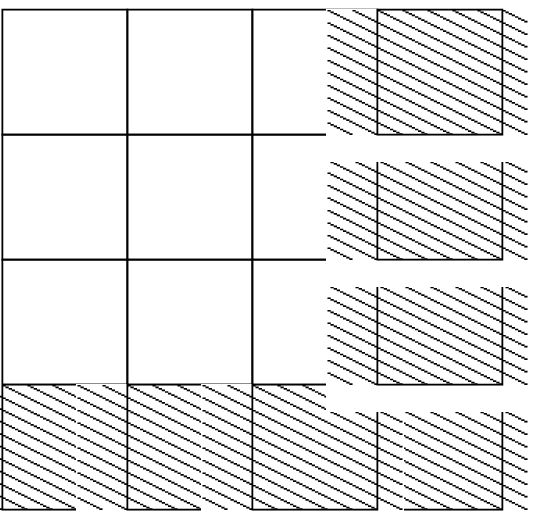}
(c) \includegraphics[scale=.3]{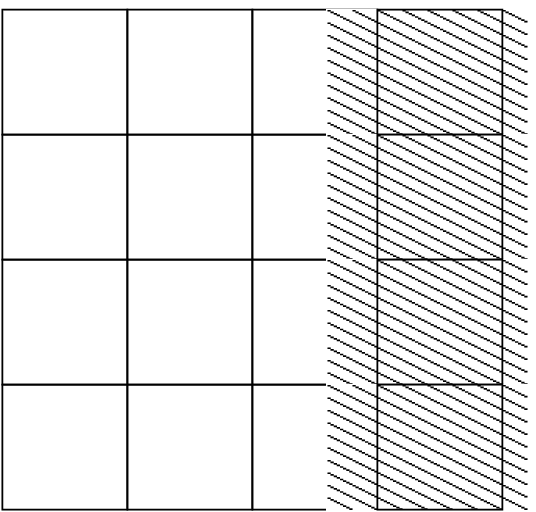}
(d) \includegraphics[scale=.3]{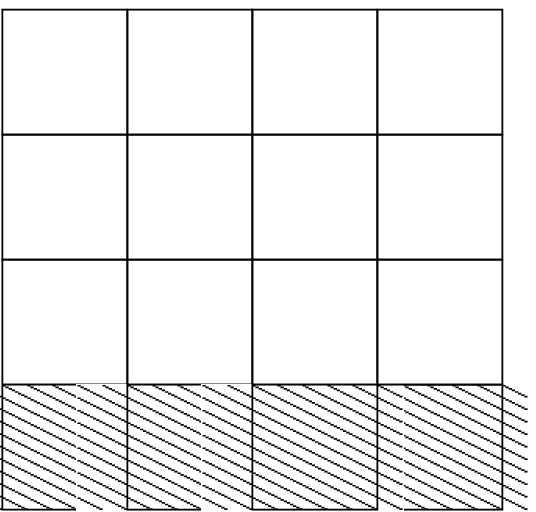}\\[2mm]
(e) \includegraphics[scale=.3]{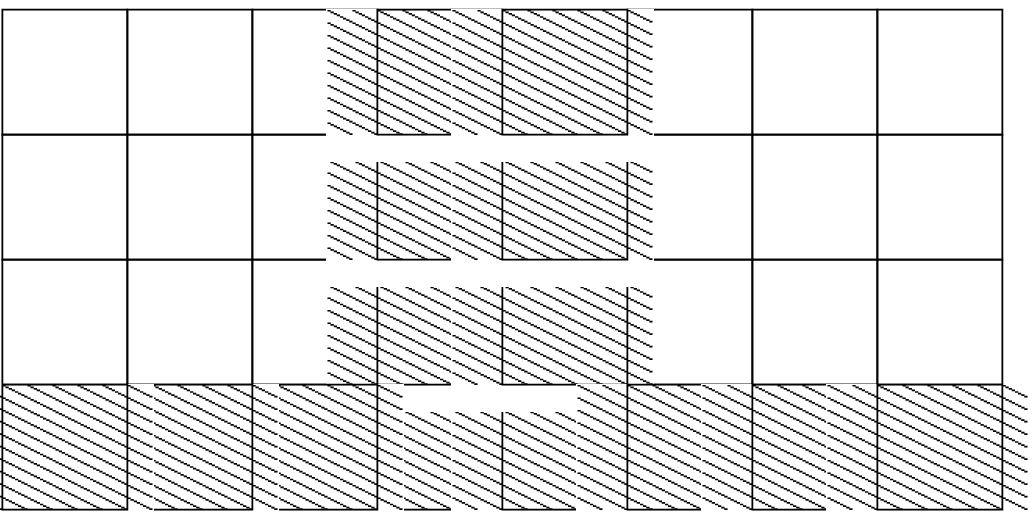}
(f) \includegraphics[scale=.3]{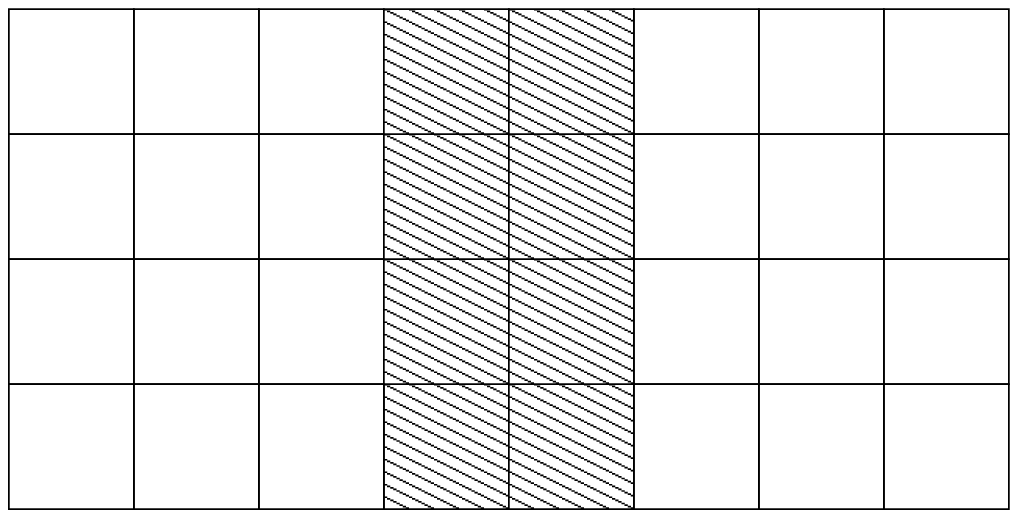}
(g) \includegraphics[scale=.3]{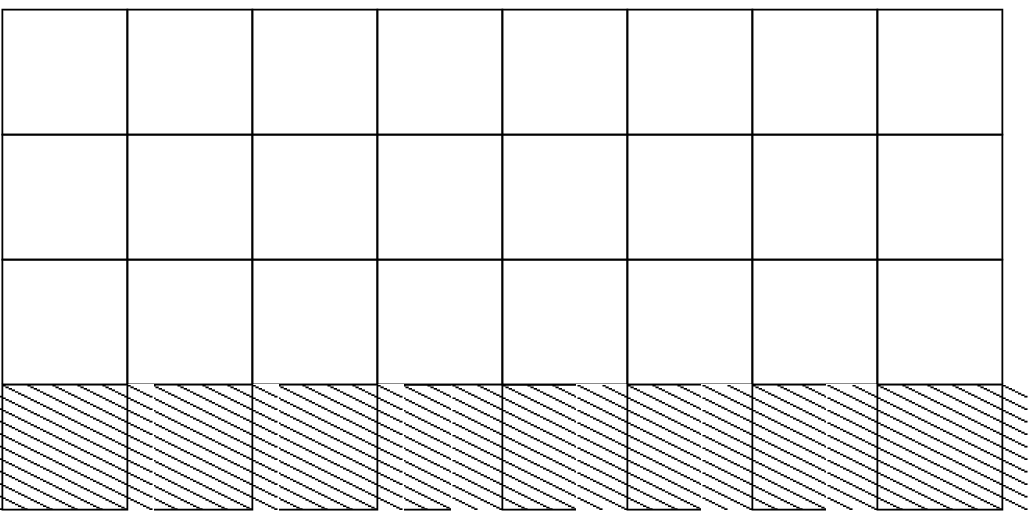}\\[2mm]
(h) \includegraphics[scale=.3]{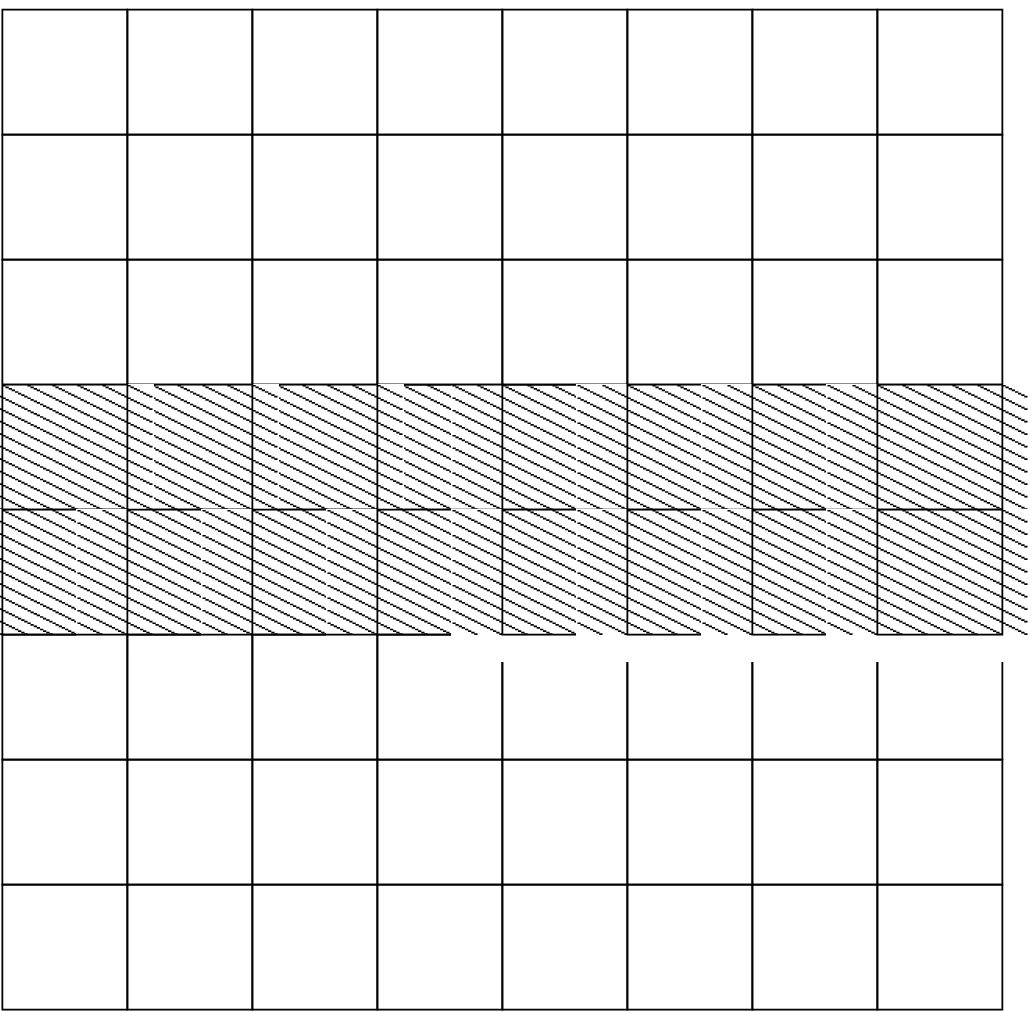}
\end{center}
\caption{
Subsequent stages of the computation of the histogram
$\Omega_{8\times 8}(n,k)$ described in Sec. \ref{sec_algorithm}.}
\label{fig_net}
\end{figure}

	In Fig. \ref{fig_net}(a) a $L^2$-long bit-string equivalent to $L\times
L$ large array (see Fig. \ref{fig_net}(b)) is presented. For $L=4$ the
bit-string corresponds to an integer number from the interval
$[-2^{15},2^{15})$. This correspondence allows to investigate all possible
lattice configurations in a simple manner. Dark sites in Fig. \ref{fig_net}(a)
correspond to the dark border of the lattice in Fig.
\ref{fig_net}(b), and they can be represented by an integer number $0\le b^7\le
127$. The first four bits of $b^7$ (marked as dark sites in Fig.
\ref{fig_net}(c)) allow to determine the additional number of anti-parallel
bonds, created by merging two $4\times 4$ lattices together to get a $8\times
4$ lattice (see Fig. \ref{fig_net}(f)). On the other hand, the last four digits
of $b^7$ (dark sites in Fig. \ref{fig_net}(d)) are equivalent to the part of the
border $0\le b^8\le 255$ (the dark sites in Fig. \ref{fig_net}(g)) of $8\times
4$ lattice.

	In the next step, two $8\times 4$ lattices are merged to create
the desired $8\times 8$ large lattice, as presented in Fig. \ref{fig_net}(h).

	Histograms ($\Omega_{8\times 4}$ and $\Omega_{8\times 8}$) for
the lattices larger than $4\times 4$ may also be easily computed basing on
$\Omega_{4\times 4}$:
\[
\label{eq_omega_8x4}
\Omega_{8\times 4}(b^8,n_1+n_2,k_1+k_2+k')=\sum_{\substack{b^7_1,n_1,k_1\\ b^7_2,n_2,k_2}}
\Omega_{4\times 4}(b^7_1,n_1,k_1)\cdot\Omega_{4\times 4}(b^7_2,n_2,k_2),
\]
where $0\le k'\le 4$ is the additional number of anti-parallel bonds in the 
darker part of Fig. \ref{fig_net}(f) and $b^8$ is combined from $b_1^7$ and
$b_2^7$.

	The histogram $\Omega_{8\times 8}(n,k)$ may be constructed in a
similar way: 
\[
\label{eq_omega_8x8}
\Omega_{8\times 8}(n_1+n_2,k_1+k_2+k'')=\sum_{\substack{b^8_1,n_1,k_1\\ b^8_2,n_2,k_2}}
\Omega_{8\times 4}(b^8_1,n_1,k_1)\cdot\Omega_{8\times 4}(b^8_2,n_2,k_2),
\]
and again $0\le k''\le 8$ is the number of anti-parallel bonds in the
darker part of Fig. \ref{fig_net}(h).

	This procedure on SGI 2800 machine takes {\em only} 22 hours of the
machine time instead of a few million years in case of the usage of the
traditional/direct method. Successive merging may be repeated
recursively to obtain the partition function for larger lattices.

\subsection{Monte Carlo simulation, mean field approach and experimental data}
To evaluate the results obtained for the $8\times 8$ lattice, we show also the
data obtained by the Monte Carlo simulations, the results of the mean field model, and
--- last but not least --- the experimental data. Standard Monte Carlo Metropolis algorithm 
\cite{herr} is applied to determine the magnetic spin
susceptibility of a $1000\times 1000$ Ising lattice. After getting
equilibrium, each point of the plot is obtained as the time average over a thousand of time steps.

	The mean field model is a direct generalization of the case of a ferromagnet.
Namely, we solve numerically a set of two equations for two sublattices
$\alpha$ and $\gamma$:
\begin{equation}
\begin{cases}
m_{\alpha }=&\tanh\big(\beta (Jm_{\gamma }+H)\big)\\
m_{\gamma }=&\tanh\big(\beta (Jm_{\alpha }+H)\big)
\end{cases}
\end{equation}
where $J<0$. In this model, the Curie temperature $T_C=-J/k_B$ and the
susceptibility $\chi$ is found as $(m_{\alpha }+m_{\gamma })/H$ for a small value of applied magnetic field, e.g. $H=0.001$.

The experimental data are collected from Ref. \cite{exp}. They concern
two-dimensional Ising antiferromagnets Rb$_2$CoF$_4$ and K$_2$CoF$_4$ where $S=\pm 1$.
The Van
Vleck susceptibility is subtracted to obtain a pure spin contribution.

\section{Results and discussion}
\begin{figure}
\begin{center}
(a) \includegraphics[angle=-90,width=.8\textwidth]{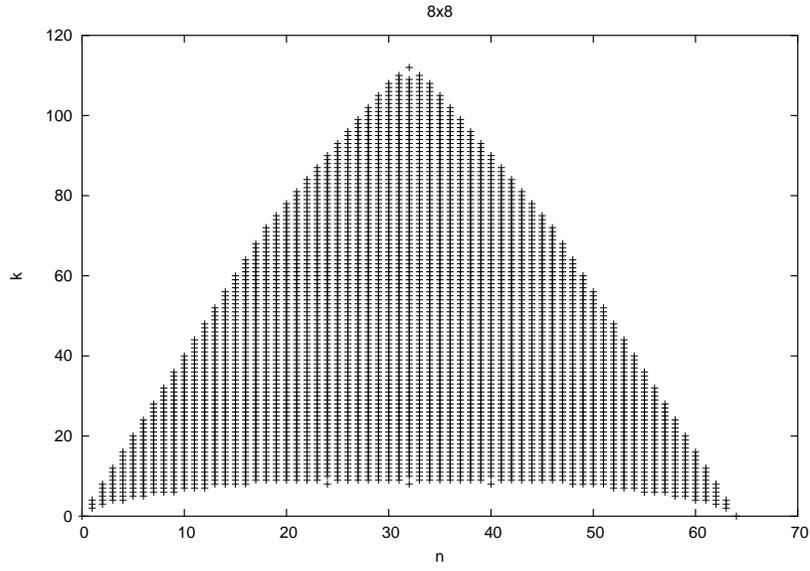}\\
(b) \includegraphics[angle=-90,width=.8\textwidth]{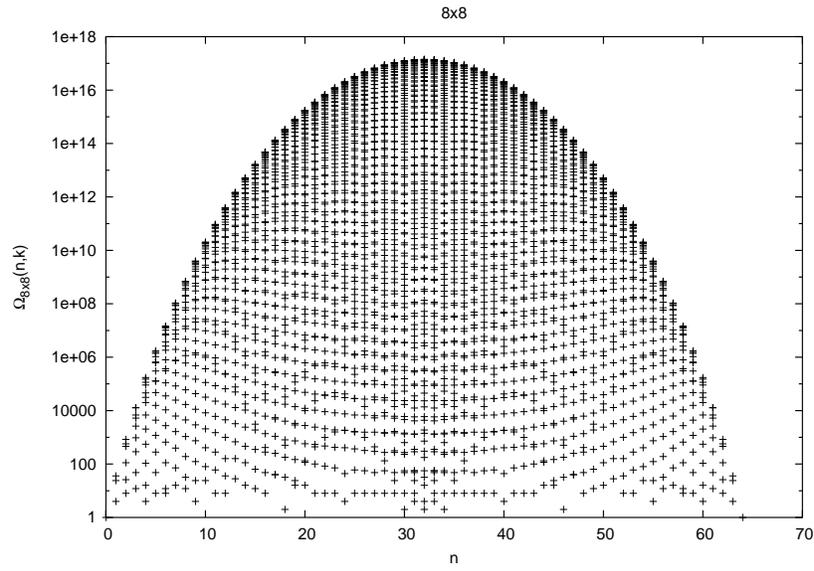}
\end{center}
\caption{
The histogram $\Omega_{8\times 8}(n,k)$.}
\label{fig_omega}
\end{figure}

In Fig. \ref{fig_omega} the histogram $\Omega_{8\times 8}(n,k)$
obtained by means of the method described in Sec. \ref{sec_algorithm} is
presented. The numbers are available at our web page \cite{omega}. The
symbols on $n$-$k$ plane (see Fig. \ref{fig_omega}(a)) indicate $(n,k)$ pairs
with non-zero values of $\Omega_{8\times 8}(n,k)$. In Fig. \ref{fig_omega}(b)
the number of configuration $\Omega_{8\times 8}(n,k)$ is presented. For fixed
$n$ different symbols correspond to different $k$. 

\begin{figure}
\begin{center}
\includegraphics[angle=-90,width=.8\textwidth]{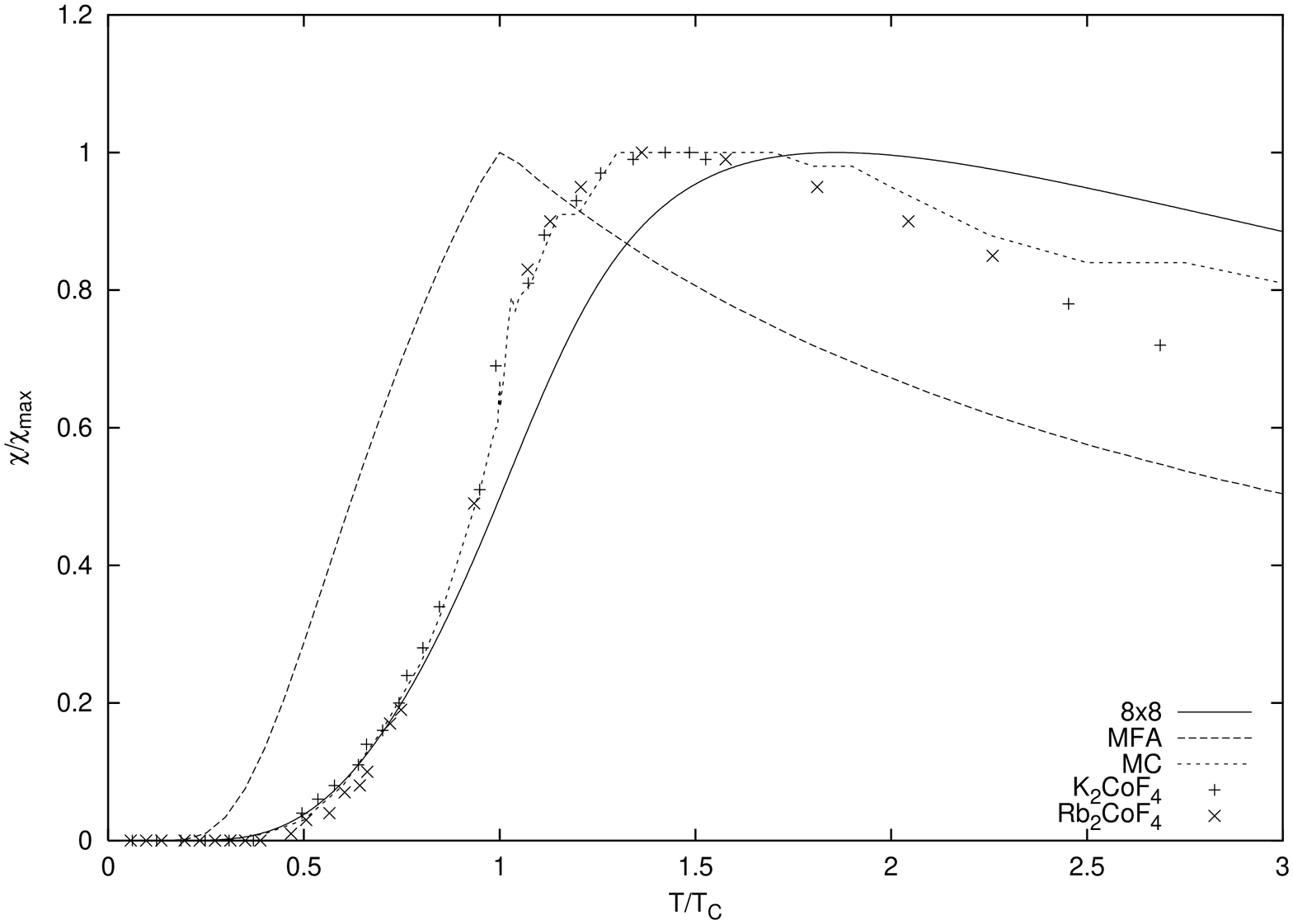}
\end{center}
\caption{
The antiferromagnetic susceptibility $\chi$ normalized to
its maximal value $\chi_{\text{max}}$, as dependent on temperature $T$
normalized to the Curie temperature $T_c$.}
\label{fig_chi}
\end{figure}

	In Fig. \ref{fig_chi} we show the thermal dependence of the antiferromagnetic
($J<0$) susceptibility, defined in Eq.~\eqref{eq_chi}.
The Curie temperature is assumed to be at the inflexion point of the curve, which is at the half of the peak height. 
Below the Curie temperature, the values of $\chi $ obtained with our
new algorithm fit well the results obtained with the Monte Carlo results and
the experimental data. Above $T_C$, the agreement is only qualititative. We hope that it
can be also quantitative if periodic boundary conditions are applied. However, in this
case the lattice border is longer, and so is the computation time. Still,
even with the present method the qualitative accordance is better than the
results of the mean field theory. 

	It would be of interest to apply the finite size scaling to our results to evaluate the critical properties of the system.
Then, the results could be compared with other methods, e.g. the finite size scaling renormalization group \cite{oliviera} or the phenomenological renormalization \cite{nightingale}.
For this purpose, however, periodic boundary conditions seem to be more appropriate as a starting point of the calculations.

	The function $\Omega (n,k)$, once known, can be easily used for the
calculation of all equilibrium thermodynamic properties, for ferro- and
antiferromagnets, various values of temperature and magnetic field. The
summation over $n$ and $k$ is much faster, than the summation over $2^{64}$
spin configurations. The results can be relevant also for other applications of
the Ising model, e.g. for the percolation problem. As for our knowledge, the
partition function has never been calculated exactly for the lattice as large
as $8\times 8$. In principle, the algorithm can be applied to larger lattices,
with a cost of more time and memory.

	The computational mountain remains
infinite, but its slope is a little bit reduced.     

\bigskip 

\noindent
{\bf Acknowledgements.}
The authors are grateful to Dietrich Stauffer for valuable comments.
The simulations were carried out in ACK-\-CY\-FRO\-NET-\-AGH.
The machine time on SGI 2800 is financed by the Polish State Committee for
Scientific Research (KBN) with grant No. KBN/\-SGI2800/\-AGH/\-077/\-2002.

\end{document}